\documentclass[12pt,preprint]{aastex}

\slugcomment{Accepted for publication by ApJ}

\begin{document}

\title{TeV gamma-ray survey of the Northern sky using the ARGO-YBJ detector}


\author{B.~Bartoli\altaffilmark{1,2},
 P.~Bernardini\altaffilmark{3,4},
 X.J.~Bi\altaffilmark{5},
 I.~Bolognino\altaffilmark{6,7},
 P.~Branchini\altaffilmark{8},
 A.~Budano\altaffilmark{8},
 A.K.~Calabrese Melcarne\altaffilmark{9},
 P.~Camarri\altaffilmark{10,11},
 Z.~Cao\altaffilmark{5},
 R.~Cardarelli\altaffilmark{11},
 S.~Catalanotti\altaffilmark{1,2},
 S.Z.~Chen\altaffilmark{0,5}\footnotetext[0]{Corresponding author: Songzhan Chen, \mbox{chensz@ihep.ac.cn}},
 T.L.~Chen\altaffilmark{12},
 Y.~Chen\altaffilmark{5},
 P.~Creti\altaffilmark{4},
 S.W.~Cui\altaffilmark{13},
 B.Z.~Dai\altaffilmark{14},
 A.~D'Amone\altaffilmark{3,4},
 Danzengluobu\altaffilmark{12},
 I.~De Mitri\altaffilmark{3,4},
 B.~D'Ettorre Piazzoli\altaffilmark{1,2},
 T.~Di Girolamo\altaffilmark{1,2},
 X.H.~Ding\altaffilmark{12},
 G.~Di Sciascio\altaffilmark{11},
 C.F.~Feng\altaffilmark{15},
 Zhaoyang Feng\altaffilmark{5},
 Zhenyong Feng\altaffilmark{16},
 Q.B.~Gou\altaffilmark{5},
 Y.Q.~Guo\altaffilmark{5},
 H.H.~He\altaffilmark{5},
 Haibing Hu\altaffilmark{12},
 Hongbo Hu\altaffilmark{5},
 Q.~Huang\altaffilmark{16},
 M.~Iacovacci\altaffilmark{1,2},
 R.~Iuppa\altaffilmark{10,11},
 H.Y.~Jia\altaffilmark{16},
 Labaciren\altaffilmark{12},
 H.J.~Li\altaffilmark{12},
 J.Y.~Li\altaffilmark{15},
 X.X.~Li\altaffilmark{5},
 G.~Liguori\altaffilmark{6,7},
 C.~Liu\altaffilmark{5},
 C.Q.~Liu\altaffilmark{14},
 J.~Liu\altaffilmark{14},
 M.Y.~Liu\altaffilmark{12},
 H.~Lu\altaffilmark{5},
 L.L.~Ma\altaffilmark{5},
 X.H.~Ma\altaffilmark{5},
 G.~Mancarella\altaffilmark{3,4},
 S.M.~Mari\altaffilmark{8,17},
 G.~Marsella\altaffilmark{3,4},
 D.~Martello\altaffilmark{3,4},
 S.~Mastroianni\altaffilmark{2},
 P.~Montini\altaffilmark{8,17},
 C.C.~Ning\altaffilmark{12},
 M.~Panareo\altaffilmark{3,4},
 B.~Panico\altaffilmark{10,11},
 L.~Perrone\altaffilmark{3,4},
 P.~Pistilli\altaffilmark{8,17},
 F.~Ruggieri\altaffilmark{8},
 P.~Salvini\altaffilmark{7},
 R.~Santonico\altaffilmark{10,11},
 S.N.~Sbano\altaffilmark{3,4},
 P.R.~Shen\altaffilmark{5},
 X.D.~Sheng\altaffilmark{5},
 F.~Shi\altaffilmark{5},
 A.~Surdo\altaffilmark{4},
 Y.H.~Tan\altaffilmark{5},
 P.~Vallania\altaffilmark{18,19},
 S.~Vernetto\altaffilmark{18,19},
 C.~Vigorito\altaffilmark{19,20},
 B.~Wang\altaffilmark{5},
 H.~Wang\altaffilmark{5},
 C.Y.~Wu\altaffilmark{5},
 H.R.~Wu\altaffilmark{5},
 B.~Xu\altaffilmark{16},
 L.~Xue\altaffilmark{15},
 Q.Y.~Yang\altaffilmark{14},
 X.C.~Yang\altaffilmark{14},
 Z.G.~Yao\altaffilmark{5},
 A.F.~Yuan\altaffilmark{12},
 M.~Zha\altaffilmark{5},
 H.M.~Zhang\altaffilmark{5},
 Jilong Zhang\altaffilmark{5},
 Jianli Zhang\altaffilmark{5},
 L.~Zhang\altaffilmark{14},
 P.~Zhang\altaffilmark{14},
 X.Y.~Zhang\altaffilmark{15},
 Y.~Zhang\altaffilmark{5},
 J.~Zhao\altaffilmark{5},
 Zhaxiciren\altaffilmark{12},
 Zhaxisangzhu\altaffilmark{12},
 X.X.~Zhou\altaffilmark{16},
 F.R.~Zhu\altaffilmark{16},
 Q.Q.~Zhu\altaffilmark{5} and
 G.~Zizzi\altaffilmark{9}\\ (The ARGO-YBJ Collaboration)}

\altaffiltext{1}{Dipartimento di Fisica dell'Universit\`a di Napoli
                  ``Federico II'', Complesso Universitario di Monte
                  Sant'Angelo, via Cinthia, 80126 Napoli, Italy.}
 \altaffiltext{2}{Istituto Nazionale di Fisica Nucleare, Sezione di
                  Napoli, Complesso Universitario di Monte
                  Sant'Angelo, via Cinthia, 80126 Napoli, Italy.}
 \altaffiltext{3}{Dipartimento Matematica e Fisica "Ennio De Giorgi",
                  Universit\`a del Salento,
                  via per Arnesano, 73100 Lecce, Italy.}
 \altaffiltext{4}{Istituto Nazionale di Fisica Nucleare, Sezione di
                  Lecce, via per Arnesano, 73100 Lecce, Italy.}
 \altaffiltext{5}{Key Laboratory of Particle Astrophysics, Institute
                  of High Energy Physics, Chinese Academy of Sciences,
                  P.O. Box 918, 100049 Beijing, P.R. China. }
 \altaffiltext{6}{Dipartimento di Fisica dell'Universit\`a di
                  Pavia, via Bassi 6, 27100 Pavia, Italy.}
 \altaffiltext{7}{Istituto Nazionale di Fisica Nucleare, Sezione di Pavia,
                  via Bassi 6, 27100 Pavia, Italy.}
 \altaffiltext{8}{Istituto Nazionale di Fisica Nucleare, Sezione di
                  Roma Tre, via della Vasca Navale 84, 00146 Roma, Italy.}
 \altaffiltext{9}{Istituto Nazionale di Fisica Nucleare - CNAF, Viale
                  Berti-Pichat 6/2, 40127 Bologna, Italy.}
 \altaffiltext{10}{Dipartimento di Fisica dell'Universit\`a di Roma ``Tor Vergata'',
                   via della Ricerca Scientifica 1, 00133 Roma, Italy.}
 \altaffiltext{11}{Istituto Nazionale di Fisica Nucleare, Sezione di
                   Roma Tor Vergata, via della Ricerca Scientifica 1,
                   00133 Roma, Italy.}
 \altaffiltext{12}{Tibet University, 850000 Lhasa, Xizang, P.R. China.}
 \altaffiltext{13}{Hebei Normal University, Shijiazhuang 050016,
                   Hebei, P.R. China.}
 \altaffiltext{14}{Yunnan University, 2 North Cuihu Rd., 650091 Kunming,
                   Yunnan, P.R. China.}
 \altaffiltext{15}{Shandong University, 250100 Jinan, Shandong, P.R. China.}
 \altaffiltext{16}{Southwest Jiaotong University, 610031 Chengdu,
                   Sichuan, P.R. China.}
 \altaffiltext{17}{Dipartimento di Fisica dell'Universit\`a ``Roma Tre'',
                   via della Vasca Navale 84, 00146 Roma, Italy.}
 \altaffiltext{18}{Osservatorio Astrofisico di Torino dell'Istituto Nazionale
                   di Astrofisica, corso Fiume 4, 10133 Torino, Italy.}
 \altaffiltext{19}{Istituto Nazionale di Fisica Nucleare,
                   Sezione di Torino, via P. Giuria 1, 10125 Torino, Italy.}
 \altaffiltext{20}{Dipartimento di Fisica dell'Universit\`a di
                   Torino, via P. Giuria 1, 10125 Torino, Italy.}

\begin{abstract}
The ARGO-YBJ detector is an extensive air shower array that has been used to monitor the northern $\gamma$-ray sky at energies above 0.3 TeV from 2007 November to 2013 January. In this paper,  we   present the results of a sky survey in the declination band from $-10^{\circ}$ to $70^{\circ}$, using  data recorded over the past five years.  With an integrated sensitivity ranging from 0.24 to $\sim$1 Crab units depending on the declination,  six  sources have been detected   with a statistical significance greater than 5 standard deviations.  Several excesses are also reported as potential $\gamma$-ray emitters. The features of each source are presented and discussed.  Additionally,  95\% confidence level upper limits of the flux from the investigated sky region  are shown. Specific upper limits for 663 GeV $\gamma$-ray AGNs inside the ARGO-YBJ field of view are reported. The effect of the absorption of $\gamma$-rays due to the interaction with   extragalactic background light is estimated.

\end{abstract}


\section{Introduction}
Over the past two decades, great advances have been made in  very high energy (VHE) $\gamma$-ray astronomy and almost 150   sources have been observed by ground-based $\gamma$-ray detectors.
Several categories of VHE $\gamma$-ray emitters
have been firmly established: Active
Galactic Nuclei (AGNs), Pulsar Wind Nebulae (PWNs), SuperNova Remnants (SNRs),  X-ray Binaries
(XBs), and starburst galaxies.
VHE $\gamma$-ray astronomy, therefore,  has progressively introduced new ways to probe the non-thermal universe and the extreme physical processes in  astrophysical sources.
VHE $\gamma$-rays are emitted by  relativistic particles  accelerated at the astrophysical shocks that are widely believed to exist in all  VHE sources.   These shocks may accelerate protons or electrons. Relativistic electrons can scatter  low energy photons to VHE levels via the Inverse Compton (IC) process, while relativistic protons would lead to  hadronic cascades and VHE $\gamma$-rays are generated by the decay of secondary $\pi^{0}$ mesons. Hence, VHE $\gamma$-ray observations are also important for understanding the origin and acceleration of cosmic rays.

VHE  $\gamma$-ray emitters  include   Galactic sources and  extragalactic sources. Most of the identified Galactic sources belong to  PWNs,  SNRs, and XBs; however, about one-third of  them are still unidentified\footnote{http://tevcat.uchicago.edu/ (Version: 3.400, as of 2013 July).}. Extragalactic sources are mainly composed of blazars, including  BL-Lac-type objects and flat spectrum radio quasars (FSRQs).  Due to interaction with   extragalactic background light (EBL), which causes a substantial reduction of the flux, VHE $\gamma$-ray observations are limited to nearby sources. The most distant VHE  source located to date is 3C 279 with a redshift value of z=0.536 \citep{albert08}.

Recent advances in the observation of VHE $\gamma$-rays are mainly attributed to the successful operation of imaging atmospheric Cherenkov telescopes (IACTs),
such as H.E.S.S., MAGIC, VERITAS, and CANGAROO, which made a majority  of the discoveries when searching for counterparts of sources observed at lower energies  \cite[for a review, see][]{aharon08b}.
To achieve an overall view of the universe in the VHE $\gamma$-ray band, an unbiased sky survey is needed, similar to that carried out by $Fermi$ and
its predecessor EGRET at GeV   energies. The two surveys detected 1873 and 271 objects,respectively, including 575 and 170 sources still unidentified \citep{nolan12,hartman99}.
The H.E.S.S. collaboration has made great progress in surveying  the Galactic plane and has revealed over 60 new VHE
$\gamma$-ray sources \citep{gast11}. However, due to their  small fields of view (FOVs) and   low duty cycles,
 IACTs are not suitable  for performing a long-term comprehensive sky survey. Although with a sensitivity lower than that of IACTs,
extensive air shower (EAS) arrays, such as   Tibet AS$\gamma$, Milagro, and ARGO-YBJ, are  the only  choices available for performing
a continuous sky survey of VHE sources. To date, several surveys   have been performed by  AIROBICC  \citep{aharon02}, Milagro   \citep{atkins04}, and Tibet AS$\gamma$   \citep{amenom05}. The latter two surveys have resulted in the successful observation of
$\gamma$-ray emissions from the Crab Nebula and Mrk 421. The best upper limits at energies above 1 TeV are around 0.27$-$0.60 Crab units  achieved by the Milagro experiment. In 2007, Milagro updated its survey of the Galactic plane and three new extended sources were discovered \citep{abdo07a}. Additionally, both Milagro and AS$\gamma$ have observed some excesses from  positions associated with the  $Fermi$  Bright Source List inside the Galactic plane \citep{abdo09,amenom10}.

The ARGO-YBJ detector is an EAS array with a large
 FOV and  can continuously monitor the  sky in the declination band from  $-$10$^{\circ}$ to 70$^{\circ}$.
With its full coverage configuration and its location at a  high  altitude of 4300 m a.s.l., the energy threshold of ARGO-YBJ is much lower than that of any previous EAS array. Since the $\gamma$-ray absorption due to EBL increases with the $\gamma$-ray energy, ARGO-YBJ, working with a threshold of a few hundred GeV, is suitable for observing AGNs that account for  80\% of the known $\gamma$-ray sources as revealed by $Fermi$\citep{nolan12}.
Previously, the ARGO-YBJ collaboration    reported the search for emission of GeV-TeV photons from GRBs \citep{aielli09b,aielli09d} and  the observation of flaring activity from  AGNs \citep{barto11,barto12b}, and   specific observations for extended sources inside the Galactic plane \citep{barto12a,barto12c,barto13}.
This paper present the analysis of a sky survey that searched for steady VHE $\gamma$-ray emitters  using more than five years of  data collected by ARGO-YBJ.

\section{The ARGO-YBJ detector}
The ARGO-YBJ detector, located at the Yangbajing Cosmic Ray Observatory (Tibet,  China, 90.5$^{\circ}$ east, 30.1$^{\circ}$ north),  is designed for VHE $\gamma$-ray astronomy and cosmic-ray observations.   It consists of a single layer of Resistive Plate Chambers (RPCs; 2.8 m $\times$1.25 m)   equipped with 10 logical pixels (called pads, 55.6 cm $\times$ 61.8 cm each) used for triggering and timing purposes.
One-hundred and thirty clusters (each composed of 12 RPCs) are installed to form the central carpet of 74 m $\times$ 78 m with an active area of $\sim$92\%, surrounded by 23 additional partially instrumented  clusters (the ``guard ring''). The total area of the array is 110 m $\times$ 100 m.
Further details about the detector and the RPC performance can be found in \cite{aielli06,aielli09c}. The arrival
time of a particle is measured with a resolution of approximately
1.8 ns. In order to calibrate the 18,360 Time to Digital
Converter  channels, we have developed a method using cosmic ray showers \citep{he07}.
The calibration precision is 0.4 ns  and the procedure is applied every month \citep{aielli09}.

The central 130 clusters began recording data in  2006 July, while the ``guard ring'' was merged into the DAQ stream in  2007 November.
The ARGO-YBJ detector is operated by requiring the  coincidence of at least 20 fired pads
(N$_{pad}$) within 420 ns on the entire carpet detector.
The time of each fired pad in a window of 2 $\mu s$   around the
trigger time and its location are recorded.
The trigger rate is 3.5 kHz with a dead time of 4\%
and the average duty-cycle is higher than $86\%$.

The high granularity of the apparatus permits a detailed space-time reconstruction of the
shower profile, including  the shower core and incident direction of the primary particle.
The shower core is estimated using a maximum likelihood method by fitting the  lateral
density distribution  of the shower with  an Nishimura-Kamata-Greisen-like function. The core resolution (68\% containment) is better than 10 m for events with    N$_{pad}>$100, and worsens for events with fewer pads.
The incident direction is reconstructed using the least squares method assuming a conical shape of the shower front.
The conical correction coefficient defined in Equation (1) of \citet[][]{aielli09}, which describes the increase of time delay with the distance to the shower core,   is fixed at  0.1 ns m$^{-1}$. According to \cite{eckma91}, a systematic  inclination of the reconstructed shower direction exists if the shower core is near the edge of the detector array.  The effect  has been confirmed using ARGO-YBJ simulation data samples and has been corrected using the method presented in  \cite{eckma91}. This correction has little effect  for events with N$_{pad}<$100 due to the large uncertainty in the core location, while it can improve the angular resolution for events with N$_{pad}>$200 by $\sim$20\%.
The improvement is better at higher multiplicities.

To improve the sensitivity for $\gamma$-ray source observation, an optimization on the selection of the shower core position is applied.
The  event selections  are listed in Table 1,
where $R$ is the distance between the shower core position and the carpet center, and TS
is the time spread of the shower front in the conical fit defined in Equation (1) of \citet[][]{aielli09}.
With these selections, more background cosmic rays  than $\gamma$-rays are rejected and  the corresponding angular resolutions are also improved.
Therefore,  the sensitivity is improved by 10\%$-$30\% with respect to that with no event selection for a Crab-like source in different $N_{pad}$ ranges. The  angular resolution ($\sigma_{res}$) for events with different multiplicities is   listed in Table 1.   The point-spread function (PSF) is fitted using a symmetrical two-dimensional Gaussian function with sigma=$\sigma_{res}$. The angular resolution listed in Table 1 is for a $\gamma$-ray shower.
The median energies  depend on both  the $\gamma$-ray spectral index and the source declination.
The median energies exhibited in Table 1 are for $\gamma$-rays from the Crab Nebula.

The effective area of the ARGO-YBJ detector for detecting
$\gamma$-ray showers is estimated   using a full Monte Carlo simulation
driven by CORSIKA 6.502 \citep{capde92} and by the GEANT4-based code
G4argo \citep{guo}. The core location of the shower is sampled inside an  area of 1000 m $\times$ 1000 m
around the carpet center. The effective areas for   $\gamma$-rays   at the three zenith angles $\theta=10^{\circ}$, $\theta=30^{\circ}$, and $\theta=50^{\circ}$ are shown in Figure 1  as a function of the primary energy   from 10 GeV to 100 TeV. The solid lines are for  all triggered events with N$_{pad}>20$.
The dotted lines show the effective areas after applying the selections listed in Table 1.
The effective area is about 100 m$^2$ at 100 GeV and $\sim$10,000 m$^2$ above 1 TeV for a zenith angle of 10$^{\circ}$.

The performance of the ARGO-YBJ detector array has been thoroughly tested by
measuring  the cosmic ray shadow cast by the Moon and    the Sun \citep{barto11b,aielli11}.
The  angular resolution obtained using the Moon shadow test is in good agreement with the Monte
Carlo simulation.
The position of the shadow
allows for the investigation of any pointing bias. The east$-$west
displacement is in good agreement with the expectation, while
a 0.1$^{\circ}$ pointing error toward the north is observed.
 By studying the westward
shift of the shadow due to the geomagnetic field, the total absolute energy
scale error, including systematic effects, is estimated to
be less than 13\% \citep{barto11b}.

\section{Data analysis}
The ARGO-YBJ data used in this analysis was  collected from 2007 November to 2013 January.
The total effective observation time is 1670.45  days. For the analysis presented in this paper, only events with  zenith angles   less than 50$^{\circ}$ are used, and   data sets  are divided into nine groups according to the number of N$_{pad}$ firing . The  event selections   listed in Table 1 are applied.
The number of events in each group and  the fraction of selected events  are also listed in Table 1. The   number of events  used in this work is 2.99$\times$10$^{11}$, which is   66.4\% of the total   number of  events recorded at zenith angles $<$50$^{\circ}$.

For the data set in each group, the sky map in celestial coordinates (right ascension and declination) is divided into a grid of
$0.1^{\circ}\times0.1^{\circ}$ bins and filled with detected events according to their
reconstructed arrival direction. The number of events is denoted as $n$. To obtain the excess of $\gamma$-induced
showers in each bin, the ``direct integral method'' \citep{fleysher04} is adopted in order to estimate the number
of cosmic ray background events in the bin, denoted as $b$. To remove the effect of cosmic ray anisotropy on a scale of $11^{\circ}\times11^{\circ}$,
a correction procedure as described in \cite{barto11} has been applied.
To reduce the contamination from the Galactic Plane diffuse $\gamma$-ray emission,
a specific similar correction procedure has been adopted in the region of Galactic latitude
$\mid b\mid <2^{\circ}$. Diffuse $\gamma$-rays are estimated on a scale of $16^{\circ}\times4^{\circ}$ in
  Galactic coordinates along the Galactic Plane,
and the contribution from a $5^{\circ}\times4^{\circ}$ window around the source bin is excluded.

In order to extract the $\gamma$-ray signals,  the events in a circular area centered on the bin
within an angular radius of 2$\sigma_{res}$  are summed after weighting
with the Gaussian-shaped PSF. Each bin is denoted as $i$. The weight is
\begin{equation}
w(r)=\frac{1}{2\pi\sigma_{res}^2}e^{-r^2/(2\sigma_{res}^2)}
\end{equation}
where $r$ is the space angle to the central bin.
Equation (9) in \cite{li83} is used to estimate the
significance of the excess in each bin. That is
\begin{equation}
S=\frac{N_{s}}{\sigma(N_s)}
\end{equation}
where
\begin{equation}
N_s=\sum_i w(r)(n_i-b_i),   \sigma(N_s)=\sqrt{\sum_i{ w^2(r)(n_i\alpha+b_i)}},
\end{equation}
The quantity $\alpha$ is the ratio of the signal and background exposures \citep{fleysher04}.
The equation above can be  used for both   one-group data sets and multi-group data sets.
For one-group data, the improvement of the significance compared to the case with $w(r)=1$ is about 10\%.
For the nine groups, the improvement is about 40\% for the analysis presented in this work, compared with the traditional method of using one average angular radius for all groups.

\section{Results}
\subsection{Sky survey results}
The pre-trial significance distribution of the bins in the whole map    is shown in Figure 2. The distribution, with a mean value of 0.002 and $\sigma=$1.02, closely follows a standard Gaussian distribution except for a tail with large positive values, due to   excesses from several $\gamma$-ray emission regions.
Figure 3  shows the significance map of the observed sky, in which  the positions of the excess regions are  visible.
Table 2 lists the locations  of   the regions with significant standard deviations (s.d.) greater than 4.5. For each independent region, only the coordinates of the pixel with the highest significance are given.
Based on the distribution of negative values (Figure 2), a significance threshold of 4.5 s.d. corresponds
to  $\sim$ 2 false sources in our catalog.

The Galactic plane is rich in potential $\gamma$-ray sources, and many VHE emitters have been detected. Recently, new candidates within the Galactic plane have been reported by Milagro and Tibet AS$\gamma$ \citep{abdo09,amenom10}. The significance distribution of the inner Galactic plane  region (longitude $20^{\circ}<l<90^{\circ}$ and latitude $\mid b\mid<2^{\circ}$) is also shown in Figure 2.  The Gaussian fit of the distribution has a mean of 0.40 and $\sigma$=1.04. In this case,   due to significant excess, a tail is present.  The locations of  the excesses  with  significance greater than 4.0 s.d. are also listed in Table 2.
The significance map of the inner Galactic plane region ( $20^{\circ}<l<90^{\circ}$, $|b|<10^{\circ}$) is shown in Figure 4.
For comparison, the   known GeV and TeV sources
 are marked in the figure.
Four regions  are significantly higher than other regions, i.e., ARGO J1839$-$0627, ARGO J1907+0627, ARGO J1912+1026  and ARGO J2031+4157.
To explore the Galactic plane  at different energies, the map obtained using events with N$_{pad}\geq100$ (corresponding to a median energy $\sim$ 1.8 TeV) is shown in the bottom panel of Figure 4.

Only pre-trial significances are reported in Table 2. It is very difficult to count
the number of trials directly, given that the significances for adjacent grid points are correlated since the smoothing  radius is larger than the grid spacing.
Since the smoothing radius is larger than the bin width, the significances in adjacent bins are correlated, and a Monte Carlo simulation is necessary to correctly evaluate the post-trial probabilities.
According to our simulations, a chance-probability less than 5\% corresponds to pre-trial significance thresholds as high as 5.1 s.d. anywhere in the map and 4.0 s.d. in the Galactic Plane.
However, since only $\sim$70 known VHE emitters exist in the sky region monitored by ARGO-YBJ, the post-trial significance increases for any candidate source associated with a counterpart.

\subsection{Characteristics of each source and source candidate}
In the following, a detailed
presentation of the sources and candidates listed in Table 2   is given.

\textbf{\emph{ARGO J0535+2203}}, detected at 21 s.d.,  is  consistent in position with the Crab Nebula.
The  location is 0.08$^{\circ}$ from the pulsar,  within the statistical error.
The spectral energy distribution (SED) derived from the ARGO-YBJ data, using the   conventional fitting method described in \cite{barto11}, in the energy range from 0.1 TeV to
35 TeV is $\frac{dN}{dE}=(3.00\pm0.18) \times 10^{-11}(E/1\; TeV)^{-2.62\pm0.06}$ (TeV$^{-1}$ cm$^{-2}$ s$^{-1}$). Only statistical errors are listed here. The integral flux of this spectrum  is denoted as I$_{crab}$ in the following text. The integral flux above 1 TeV is 1.85$\times10^{-11}$ cm$^{-2}$ s$^{-1}$. It is 5.69$\times 10^{-11}$ cm$^{-2}$ s$^{-1}$   above 500 GeV.  This SED is consistent, within the errors, with the  results obtained
by other experiments, e.g. HEGRA, H.E.S.S., MAGIC, and Tibet AS$\gamma$ \citep{aharon04,aharon06,albert08b,amenom09}. A comparison among different experiments is shown in Figure 5. The figure  shows only statistical errors. The systematic errors on the flux for point sources have been described in \cite{barto12a} and are found to be less than 30\%.  As a standard candle, the Crab Nebula is   used to estimate the sensitivity of an experiment.
The 5 s.d. one-year sensitivity and the integrated sensitivity of ARGO-YBJ are shown in Figure 6.
Events with N$_{pad}\geq20$, N$_{pad}\geq40$ and so on, are used for this estimation.  The  integrated  sensitivity using events with N$_{pad}\geq20$  is 24\% I$_{crab}$ and the corresponding one-year sensitivity is 55\% I$_{crab}$. The sensitivity  decreases as energy increases. The  integrated  sensitivity is about 1 I$_{crab}$ above an energy   of 20 TeV.

\textbf{\emph{ARGO J1105+3821}}, detected at 14 s.d.,  is   consistent in position with the blazar Mrk 421. This is an active source and many outbursts have been detected by ARGO-YBJ over the past five years \citep{aielli10,barto11,chen13}. Its five-year average SED in the energy range from 0.1 TeV to
11 TeV is $\frac{dN}{dE}=(1.35\pm0.12) \times 10^{-11}(E/1\; TeV)^{-2.75\pm0.09}$ (TeV$^{-1}$ cm$^{-2}$ s$^{-1}$). The integral flux above 1 TeV is (1.30
$\pm$ 0.11) $\times 10^{-11}$ cm$^{-2}$ s$^{-1}$, corresponding to $\sim$0.70 $I_{Crab}$.

\textbf{\emph{ARGO J1654+3945}}, detected at 9 s.d.,  is   consistent in position with the blazar Mrk 501. This source entered into an active phase in 2011 October, according to ARGO-YBJ observations \citep{barto12b}. Its five-year average SED in the energy range from 0.2 TeV to
12 TeV is $\frac{dN}{dE}=(1.01\pm0.11) \times 10^{-11}(E/1\; TeV)^{-2.37\pm0.18}$ (TeV$^{-1}$ cm$^{-2}$ s$^{-1}$). The integral flux above 1 TeV is (0.95
$\pm$ 0.10) $\times 10^{-11}$ cm$^{-2}$ s$^{-1}$, corresponding to $\sim$0.51 $I_{Crab}$.

\textbf{\emph{ARGO J1839$-$0627}} is an extended source. Most of the excess  overlaps  the extended region of the unidentified source HESS
J1841$-$055 even if the peak position is slightly displaced from the center of HESS J1841$-$055 \citep{aharon08}.
The morphology detected by H.E.S.S. exhibits a highly extended, possibly
two- or three-peaked region.
A similar  morphology is also detected by ARGO-YBJ using events N$_{pad}>100$  as shown in Figure 4.
Parameterizing the source shape with a two-dimensional Gaussian function, the extension is estimated to be $\sigma=(0.40^{+0.32}_{-0.22})^{\circ}$, which is consistent with
the H.E.S.S. measurement.
The flux  measured   by ARGO-YBJ   is   higher than that determined by H.E.S.S. by a factor of $\sim$ 3.  A detailed discussion about this object can be found in  \cite{barto13}.  Recently, a young  energetic $\gamma$-ray pulsar PSR J1838$-$0537 has been detected within its extended region \citep{plets12}. The inferred energetics suggests that HESS J1841$-$055 may contain a pulsar wind nebula powered by the pulsar.

\textbf{\emph{ARGO J1907+0627}} is closely connected to \textbf{\emph{ARGO J1910+0720}}.
ARGO J1907+0627 is   consistent in position with HESS J1908+063 \citep{aharon09}, while ARGO J1910+0720 is completely outside the extended region of   HESS J1908+063. In a previous work, these two sources have been considered as a single unique source, identified as the extended source MGRO J1908+06 with  an extension of $\sigma=0.49^{\circ} \pm0.22^{\circ}$ \citep{barto12c}. The flux determined by ARGO-YBJ was consistent with that of Milagro but higher than that of HESS by a factor of $\sim$ 3. Its extended size is also marginally larger than the H.E.S.S. result.
Therefore, MGRO J1908+06 could be a blend of the two sources.
ARGO J1907+0627 is  consistent in position with the pulsar PSR J1907+0602, and could be the associated pulsar wind nebula. Very close to ARGO J1910+0720,
a counterpart in the hard X-ray band, SWIFT J1910.8+0739(4U 1909+07)  (R.A.=287.699$^{\circ}$, Dec.=7.598$^{\circ}$ in J2000 epoch)\citep{tuell10},  is located. This X-ray source is  a high-mass X-ray binary (HMXB),  a type of source  identified as a VHE $\gamma$-ray emitter.  ARGO J1910+0720 is detected at only 4.3 s.d., and the    nearby source  ARGO J1907+0627 could contribute to the observed excess. With the current statistics we  cannot exclude the possibility of a  background fluctuation.
However, this is an interesting region for follow-up observations with  more sensitive instruments.

\textbf{\emph{ARGO J1912+1026}}, detected at 4.2 s.d., is   consistent in position with HESS J1912+101 \citep{aharon08c}. HESS J1912+101 is an extended source with an intrinsic Gaussian width  $0.26^{\circ} \pm 0.03^{\circ}$ assuming a symmetrical two-dimensional Gaussian shape.
Assuming  a power-law spectrum, the spectral index obtained by ARGO-YBJ is $-$2.68$\pm$0.35, which is   consistent with $-$2.7$\pm$0.2  obtained by H.E.S.S..  However, the flux above 1 TeV is 23\% $I_{Crab}$, much higher than the value of 9\% $I_{Crab}$  determined by H.E.S.S..
We reported a similar disagreement for the source HESS J1841-055 and MGRO J1908+06.
Further discussion for such a discrepancy can be found in \cite{barto13}.

\textbf{\emph{ARGO J2021+4038}}, in the Cygnus region,  is  consistent in position  with VER J2019+407 \citep{aliu13}, whose flux is only 3.7\% $I_{Crab}$, but the nearby extended source   ARGO J2031+4157 could contribute to   most of the excess, as shown in Figure 4.

\textbf{\emph{ARGO J2031+4157}} is a highly extended source located in the Cygnus region,    consistent in position with MGRO J2031+41 and TeV J2032+4130.
The intrinsic extension estimated using ARGO-YBJ data is $\sigma=(0.2^{+0.4}_{-0.2})^{\circ}$ \citep{barto12a}.
In this case, the measured flux is also higher than that measured by IACTs, but with a discrepancy of more than a factor 10. A detailed report on this region can  be found in \cite{barto12a}. This region is also positionally consistent with the cocoon of freshly accelerated cosmic rays detected by $Fermi$ \citep{acker11b}.

\textbf{\emph{ARGO J0409$-$0627}},   detected at 4.8 s.d., is outside the Galactic plane. No counterpart at lower energies, including GeV $\gamma$-ray and X-ray bands, has previously been  found.   Its post-trial significance is the lowest   among the sources listed in Table 2 and is less than 3 s.d..

\textbf{\emph{ARGO J1841$-$0332}} is detected at 3.4 s.d. using events N$_{pad}\geq20$ and at 4.2 s.d. using events N$_{pad}\geq100$.
This source  is observed at high zenith angles, where large systematic pointing errors are expected, therefore, it is  likely coincident with the VHE  $\gamma$-ray source HESS J1843$-$033, even though it is displaced by 0.7$^{\circ}$.
Five other   GeV $\gamma$-ray sources   surround this region, as shown in Figure 4. An observation with improved sensitivity is necessary to clarify this possible TeV emission.

\subsection{Sky upper limits}
Excluding the sources listed   in Table 2, we can set upper limits to the $\gamma$-ray flux from all the directions in the observed sky region.

To estimate the   response of the ARGO-YBJ detector,
we simulate a source located at different declinations, with a power-law spectrum in the energy range from 10 GeV to 100 TeV.
Each source is traced by means of a complete transit, i.e., 24 hr of observation.
Figure 7 shows the median energy of all $\gamma$-induced showers that trigger ARGO-YBJ, i.e., N$_{pad}\geq20$, and satisfy the event selections  for sources with different  spectral indices. When the index is $-$2.6,   similar to that of the Crab Nebula, the median energy varies from 0.64 TeV at Dec.=30$^{\circ}$ (the latitude of ARGO-YBJ) to 2.4 TeV at Dec.=$-$10$^{\circ}$ and Dec.=70$^{\circ}$. For sources with a hard spectral index $-$2.0, the corresponding range of median energy is from 1.5 TeV to 5.6 TeV. The median energy varies from  0.36 TeV to 1.1 TeV for sources with a soft spectral index $-$3.0.

The statistical method given in \cite{helen83} is used to calculate the upper limit on the number of signal events at 95\% C.L. in each bin. The number of events is transformed  into a flux using the results of the simulation.
The upper limits to the flux of $\gamma$-rays with energies above 500 GeV  for each bin are shown in Figure 8. The spectral index is assumed to be $-$2.6. The average upper limits, as a function of the declination, are shown in Figure 9.  The limits range  between  9\% and  44\% $I_{Crab}$ and
are the lowest obtained so far. The lowest limit for a spectral index $-$2.0 ($-$3.0) is 5\% (9\%) $I_{Crab}$, as shown in Figure 9.

The flux upper limits shown in Figures 8 and 9 are for point sources. For  extended sources the corresponding   flux upper limit will increase.
 For a symmetrical two-dimensional Gaussian shape with $\sigma$=0.2$^{\circ}$, the upper limit will increase by  10\%. If $\sigma$=0.3$^{\circ}$ and $\sigma$=0.5$^{\circ}$, the increase will be of 20\% and 44\%, respectively. For this estimation, we assumed a spectral index $-$2.6.

With an energy  threshold lower than any other previous EAS array, ARGO-YBJ is  suitable for  the observation of AGNs, the dominant $\gamma$-ray extragalactic sources. For an extragalactic source, the absorption of $\gamma$-rays due to
the interaction with the EBL must be taken into account.
By choosing the model proposed in \cite{franc08}, the effect of   EBL absorption on the upper limits has been evaluated, and the absorption factors with respect to a source with redshift $z=0$ are shown in Figure 10, for a source spectral index of $-$2.6. Curves for redshift values of   0.03, 0.06, 0.1, and 0.3 are shown. The flux upper limits shown in Figure 8, multiplied by the absorption  factor shown in Figure 10, give the unabsorbed flux upper limit at the source. The values of the absorption factors are about 1.5$-$2.2 for sources with a redshift $z$=0.03, and increase by a factor of $\sim$10 for sources at   $z$=0.3. The absorption is stronger (weaker) for sources with harder (softer) spectra.     Figure 11 shows   examples of the absorption factors for sources   with
spectral indices of $-$2 and $-$3.

According to the $Fermi$-LAT second AGN  catalog (2LAC), 663 AGNs are within the ARGO-YBJ FOV \citep{acker11}. Figure 12 shows the comparison of ARGO-YBJ flux upper limits with the fluxes obtained by  extrapolating to TeV energies the SEDs  measured by $Fermi$-LAT in the range 1$-$100 GeV. The extrapolation is performed assuming that the spectral index steepens by 0.5 at 100 GeV.   This spectral behavior is
physically motivated because radiative cooling is expected
to modify the electron power-law index by 1 and the  corresponding    $\gamma$-ray index by 0.5.
For convenience, we show in Figure 12 the differential fluxes at 1 TeV.
As can be seen, for 135 AGNs out of the total 663, the calculated upper limits are lower than the extrapolated fluxes, suggesting steeper spectra above
100 GeV.
Such an effect could be due to  the absorption of photons by the EBL, since the average redshift is 0.27 for BL Lac objects and 1.12 for FSRQs \citep{acker11}. As evident from Figures 10 and 11, the absorption factors are very high.
The redshift has been measured for 68 AGNs out of 135. Figure 13 shows the upper limits taking into account the EBL absorption. For 10 sources out of 68, the limits set in this work constrain the intrinsic spectra to have steeper slopes.
These AGNs are listed in Table 3, which   also reports the index measured by $Fermi$ in the range 1$-$100GeV, the differential flux extrapolated to 1 TeV, and the flux upper limits corrected for the absorption. Note that Mrk 421 and Mrk 501  have been significantly detected by ARGO-YBJ, and they are the two brightest  AGNs. Five AGNs out of eight have been detected by IACTs as VHE $\gamma$-ray sources, and the spectra are consistent with the upper limits obtained here.

The upper limits obtained here for AGNs represent the five-year averaged flux. It is well known that many AGNs exhibit strong variability (up to a factor of 10) on different timescales. The upper limits for short periods are beyond the scope of this paper.

\section{Summary}
This paper has presented the most sensitive survey to date of the sky in the declination band from $-$10$^{\circ}$ to 70$^{\circ}$ obtained with five years of ARGO-YBJ data. With an integrated sensitivity ranging from 0.24 to $\sim$1 Crab flux, depending on the declination, six sources have been observed  with a statistical significance greater than 5 s.d..
These sources  are   associated with well known TeV $\gamma$-ray emitters. Evidence for possible TeV emission from five directions is also reported. Two of these five excesses are not associated with any known counterpart and thus are potentially new TeV emitters.
Of particular interest is the candidate source ARGO J1910+0720, which is coincident in position with a HMXB.
The 95\% C.L. upper limit to the $\gamma$-ray flux from all the directions in the mentioned declination band are also reported. The integral flux limits above 500 GeV vary from 0.09 to 0.44 Crab units for a Crab-like source, depending on the declination.
The limits set by ARGO-YBJ in this work are the lowest available to date.
Specific upper limits for 663 GeV AGNs are also presented and 8 AGNs are found with intrinsic spectra steeper than expected.

\acknowledgments
 This work is supported in China by NSFC (Nos.10120130794 and 11205165),
the Chinese Ministry of Science and Technology, the
Chinese Academy of Sciences, the Key Laboratory of Particle
Astrophysics, CAS, and in Italy by the Istituto Nazionale di Fisica
Nucleare (INFN).

We also acknowledge the essential support of W.Y. Chen, G. Yang,
X.F. Yuan, C.Y. Zhao, R. Assiro, B. Biondo, S. Bricola, F. Budano,
A. Corvaglia, B. D'Aquino, R. Esposito, A. Innocente, A. Mangano,
E. Pastori, C. Pinto, E. Reali, F. Taurino and A. Zerbini, in the
installation, debugging, and maintenance of the detector.


\clearpage

\begin{deluxetable}{ccccccccc} 
\tablecolumns{7} \tablewidth{0pc} \tablecaption{Event Selections and   Number of Events}
\tablehead{\colhead{$N_{pad}$ range} & \colhead{$R$ }&\colhead{TS  } & \colhead{$\sigma_{res}$ }& Median energy&\colhead{Number of Events} &Surviving Fraction\\
& (m) & (ns$^2$) & (deg)& (TeV) &($\times10^9$) & (\%)}
\startdata
$[20,40]$ & No cut&  $<$80  &1.66 &0.36  & 128   &73.0 \\
$[40,60]$ & No cut& $<$80  & 1.34 &0.56  & 102   &74.2\\
$[60,100]$ & $<$90 & $<$80 & 0.94 &0.89  & 39.3   &53.4\\
$[100,130]$& $<$70 & $<$80 &0.71  &1.1   & 8.87   &45.1 \\
$[130,200]$& $<$65 & $<$80  &0.58 &1.4   & 8.62  &43.9 \\
$[200,500]$& $<$60 & $<$80  & 0.42 &2.8  & 8.06   &45.9 \\
$[500,1000]$& $<$50 & $<$80  & 0.31&4.5  &2.19  &48.8 \\
$[1000,2000]$& $<$40 & $<$80  &0.22&8.9  & 0.806 &45.5 \\
$[>2000]$& $<$30 & $<$80  & 0.17   &18  &0.317  &34.7\\
\enddata
\end{deluxetable}

\begin{deluxetable}{cccccccc} 
\tablecolumns{7} \tablewidth{0pc} \tablecaption{Location of the excess regions}
\tablehead{ ARGO-YBJ Name & R.A.$^a$  &  Dec.$^a$  &   l   & b  &  S   &  Associated   \\
 &   (deg)  &  (deg)  &  (deg)  &   (deg)   &   (s.d.)  &   TeV  Source  }
\startdata
ARGO J0409$-$0627&62.35  &   $-$6.45 &198.51 & $-$38.73&  4.8 & \\
ARGO J0535+2203&83.75  & 22.05 &184.59&$-$5.67& 20.8  & Crab Nebula \\
ARGO J1105+3821&166.25  & 38.35   &179.43 &65.09&14.1 & Mrk 421 \\
ARGO J1654+3945&253.55  & 39.75   &63.59&38.80& 9.4 & Mrk 501 \\
ARGO J1839$-$0627&279.95  & $-$6.45   &25.87&$-$0.36& 6.0 & HESS J1841$-$055\\
ARGO J1907+0627&286.95  & 6.45 &40.53&$-$0.68& 5.3  &HESS J1908+063 \\
ARGO J1910+0720&287.65  & 7.35  &41.65&$-$0.88& 4.3  & \\
ARGO J1912+1026&288.05  & 10.45 &44.59&0.20& 4.2  &HESS J1912+101\\
ARGO J2021+4038&305.25  & 40.65 &78.34&2.28& 4.3  & VER J2019+407 \\
ARGO J2031+4157&307.95    & 41.95   &80.58&1.38& 6.1    &MGRO J2031+41\\
 & & & & & & TeV J2032+4130 \\
\hline
ARGO J1841$-$0332&280.25    & $-$3.55  &28.58&0.70& 4.2     & HESS J1843$-$033 \\
\hline
\multicolumn{7}{l}{　Notes. }\\
\multicolumn{7}{l}{　a) R.A. and Dec. are celestial coordinates in J2000 epoch. }
\enddata
\end{deluxetable}

\begin{deluxetable}{ccccccccc} 
\tablecolumns{8} \tablewidth{0pc} \tablecaption{ARGO-YBJ upper limits for  sources in the 2LAC}
\tablehead{ Name & Associated &  R.A.$^a$  & Dec.$^a$   &  $z$  &  Index$^b$ &   Flux$^c$ &  Upper limit$^d$ &   S  \\
 (2FGL)     & TeV source &(deg) &   (deg) &    &   &    &    & (s.d.) }
\startdata
J0319.8+4130 & 	NGC 1275   &    49.950 &    41.512 &     0.018 &     2.00 & 1.95e$-$11 & 5.31e$-$12& 1.4 \\
J1015.1+4925 & 1ES 1011+496&  153.767 &    49.434 &     0.212 &     1.72 & 3.96e$-$11 & 3.23e$-$11 &$-$0.5 \\
J1104.4+3812 & Mrk 421      &  166.114 &    38.209 &     0.031 &     1.77 & 1.15e$-$10$^e$ & 2.82e$-$11 &13.9 \\ 
J1117.2+2013 &             &  169.276 &    20.235 &     0.138 &     1.70 & 1.07e$-$11 & 8.77e$-$12 &$-$1.8 \\
J1428.6+4240 & H 1426+428  &  217.135 &    42.673 &     0.129 &     1.32 & 3.49e$-$11 & 1.72e$-$11 &0.2 \\
J1653.9+3945 &  Mrk 501     &  253.468 &    39.760 &     0.034 &     1.74 & 4.09e$-$11$^f$ & 2.02e$-$11 &9.1 \\
J1744.1+1934 &1ES 1741+196 &  265.991 &    19.586 &     0.083 &     1.62 & 5.82e$-$12 & 3.99e$-$12 &$-$2.0 \\
J2039.6+5218 &             &  309.848 &    52.331 &     0.053 &     1.50 & 6.95e$-$12 & 4.25e$-$12 & $-$1.1 \\
J2323.8+4212 &             & 350.967 &    42.183 &     0.059 &     1.88 & 5.09e$-$12 & 4.42e$-$12 &$-$0.7 \\
J2347.0+5142 & 1ES 2344+514&  356.771 &    51.705 &     0.044 &     1.72 & 8.20e$-$12 & 4.50e$-$12 & $-$0.8 \\
\hline
\multicolumn{9}{l}{　Notes }\\
\multicolumn{9}{l}{　a) R.A. and Dec. are celestial coordinates in J2000 epoch quoted in the 2LAC} \\
\multicolumn{9}{l}{　     \citep{acker11}.}\\
\multicolumn{9}{l}{　b) The power-law spectral index reported in the 2LAC  \citep{acker11}.}\\
\multicolumn{9}{l}{　c) Extrapolated differential flux at 1 TeV in units of TeV$^{-1}$ cm$^{-2}$ s$^{-1}$ based on 2LAC} \\
\multicolumn{9}{l}{　     parameters \citep{acker11}.}\\
\multicolumn{9}{l}{　d) 95\% C.L. flux upper limits   at 1 TeV in units of TeV$^{-1}$ cm$^{-2}$ s$^{-1}$. }\\
\multicolumn{9}{l}{　e) The measured flux is 1.35e$-$11 and the corresponding un-absorbed flux is 2.07e$-$11. }\\
\multicolumn{9}{l}{　f) The  measured flux is 1.01e$-$11 and the corresponding un-absorbed flux is 1.61e$-$11. }
\enddata

\end{deluxetable}

\clearpage

\begin{figure}
\plotone{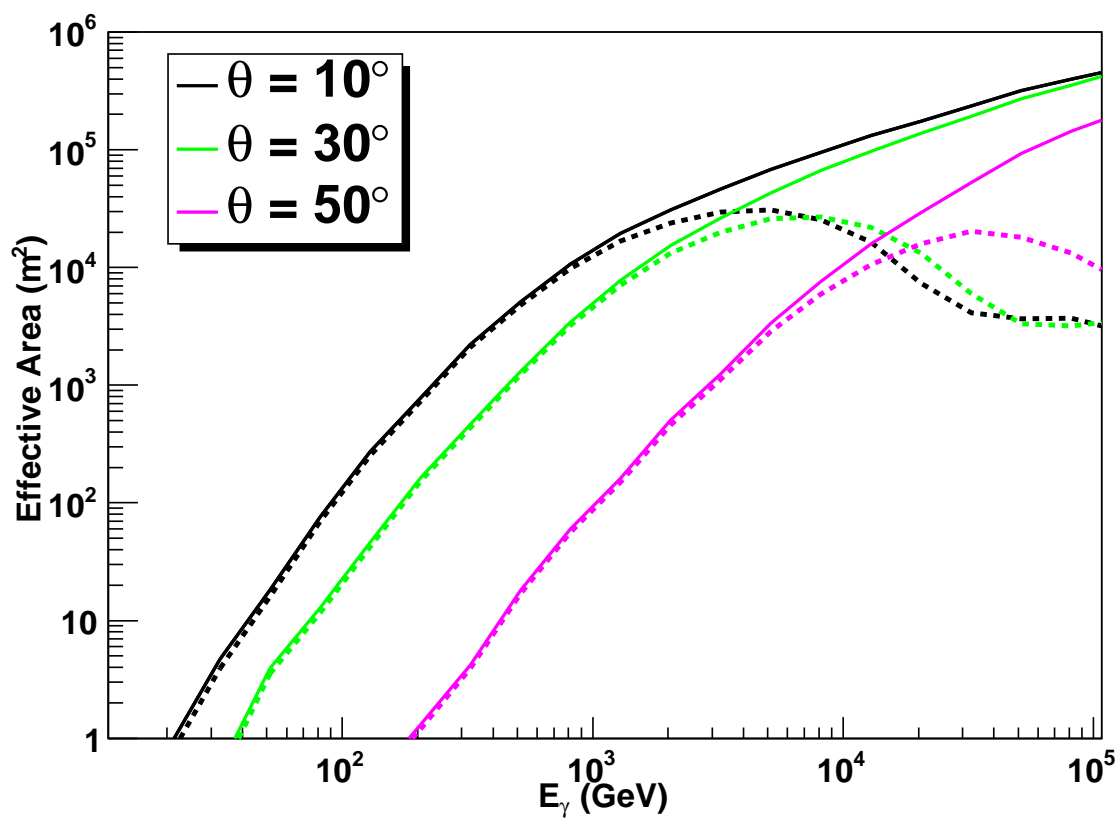}
\caption{
 ARGO-YBJ effective areas for $\gamma$-rays   as a function of
the energy for the three zenith angles $\theta=10^{\circ}$, $\theta=30^{\circ}$, and $\theta=50^{\circ}$.
The solid lines are obtained with all the triggered events (N$_{pad}\geq20$), while the dotted lines with the selected events as listed in Table 1. }
\label{fig1}
\end{figure}

\begin{figure}
\plotone{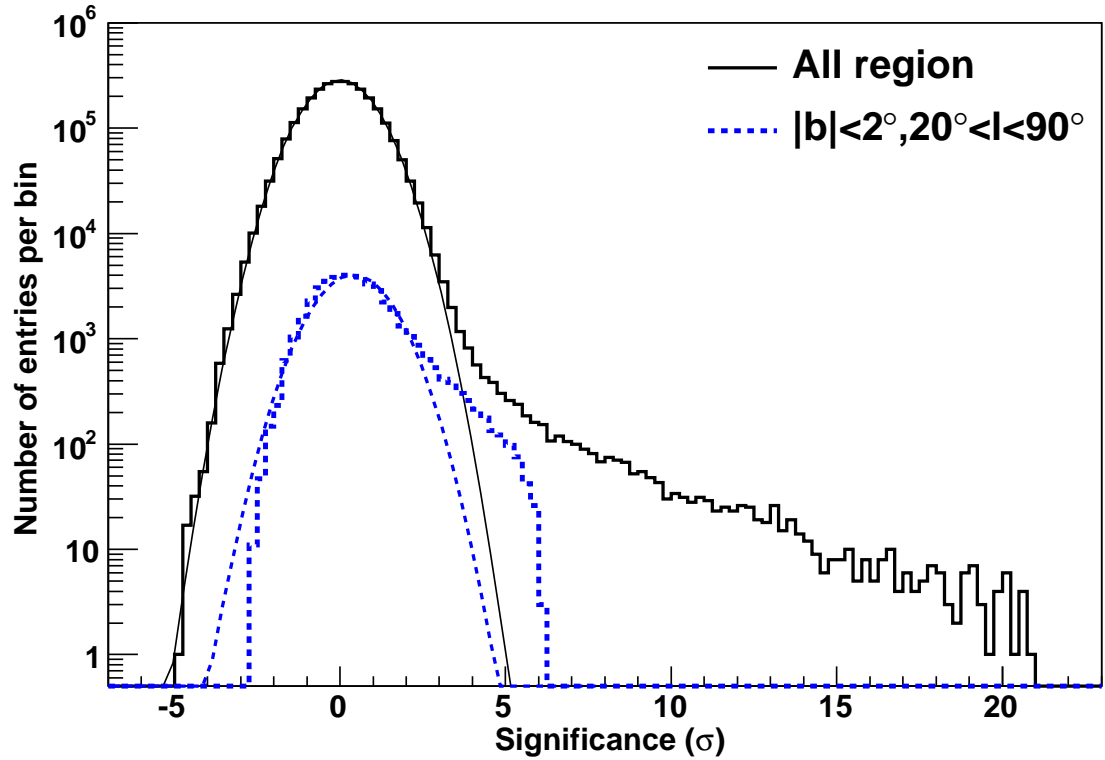}
\caption{ Pre-trial significance distribution for the whole sky map (thick solid line).
The thin solid line represents the best Gaussian fit.  The significance distribution for the Galactic Plane region with $\mid b\mid<2^{\circ}$ and $20^{\circ}<l<90^{\circ}$ is shown by the thick dotted line. The thin dotted  line represents the best Gaussian fit for this region.  }
\label{fig2}
\end{figure}

\begin{figure}
\plotone{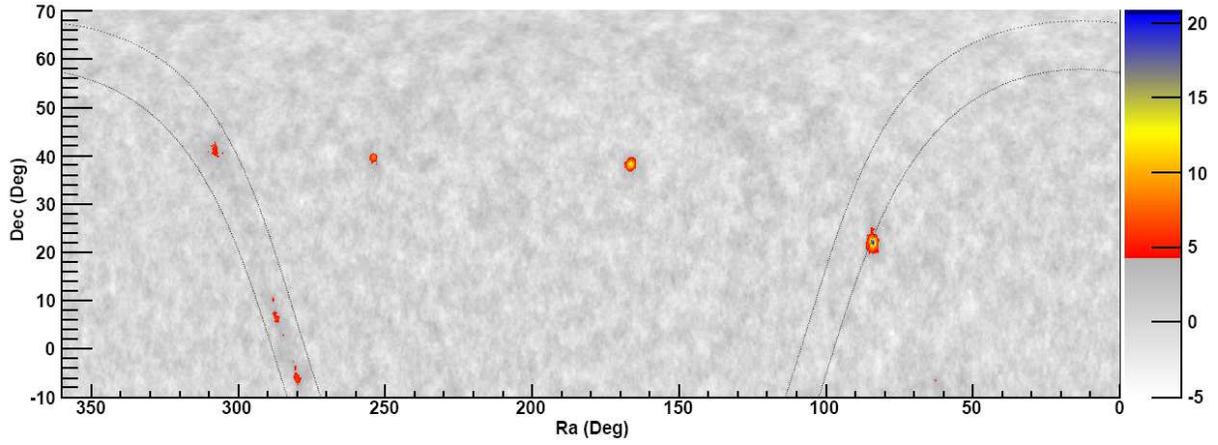}
\caption{Significance map of the   sky as seen by   ARGO-YBJ   in VHE   band. The significances of the excesses, in terms of standard deviations, are shown  by the color scale on the right side. The two dotted lines indicate the Galactic latitudes $b=\pm5^{\circ}$.  }
\label{fig3}
\end{figure}

\begin{figure}
\plotone{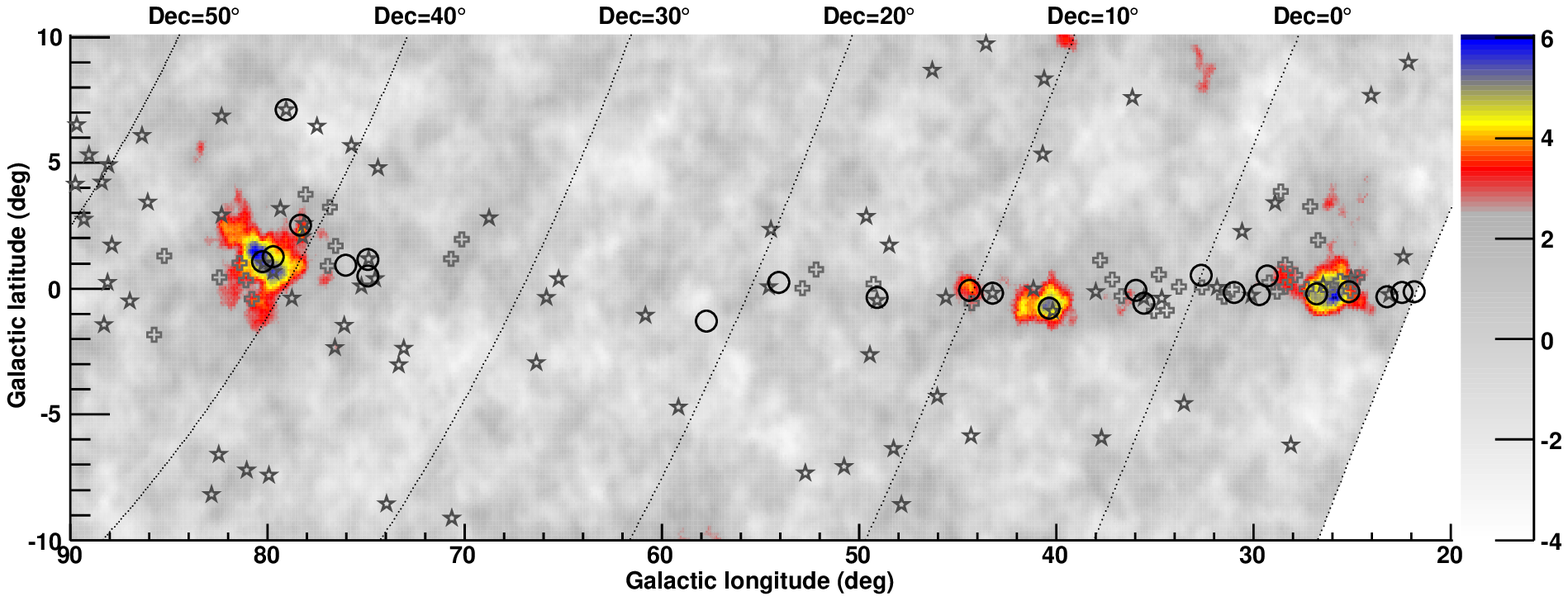}
\plotone{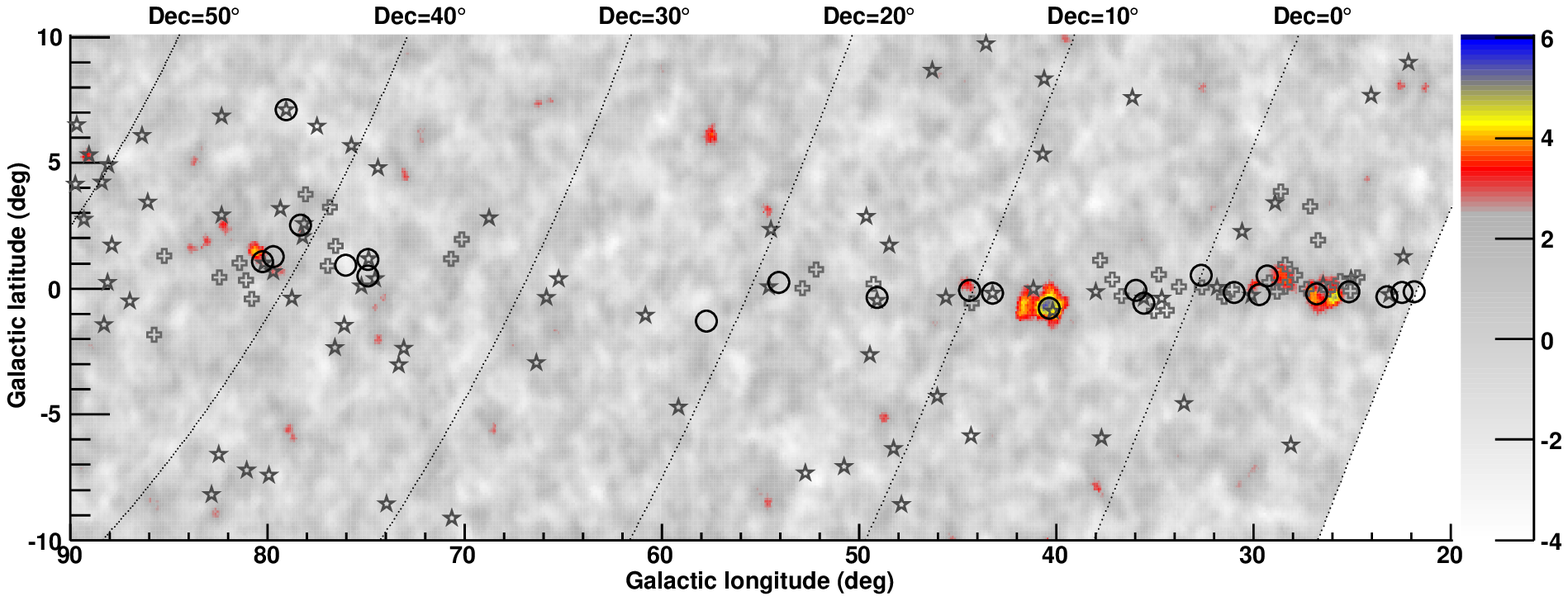}
\caption{
Significance map of the Galactic Plane  region with $\mid b\mid<10^{\circ}$ and $20^{\circ}<l<90^{\circ}$ obtained by the ARGO-YBJ
detector.  The circles indicate the positions of all the known VHE sources.  The
open stars mark the locations of the  GeV sources in the second $Fermi$-LAT   catalog \citep{nolan12}.
The open crosses mark the locations of the sources  considered to be potentially confused with Galactic diffuse emission in the second $Fermi$-LAT catalog \citep{nolan12}.
The top panel was obtained using ARGO-YBJ events with N$_{pad}\geq20$ (corresponding to a median energy $\sim$ 0.7 TeV) while the bottom panel was obtained using events with N$_{pad}\geq100$ (corresponding to a median energy $\sim$ 1.8 TeV).
The four excess regions are  ARGO J1839$-$0627, ARGO J1907+0627, ARGO J1912+1026, and ARGO J2031+4157.
}
\label{fig4}
\end{figure}

\begin{figure}
\plotone{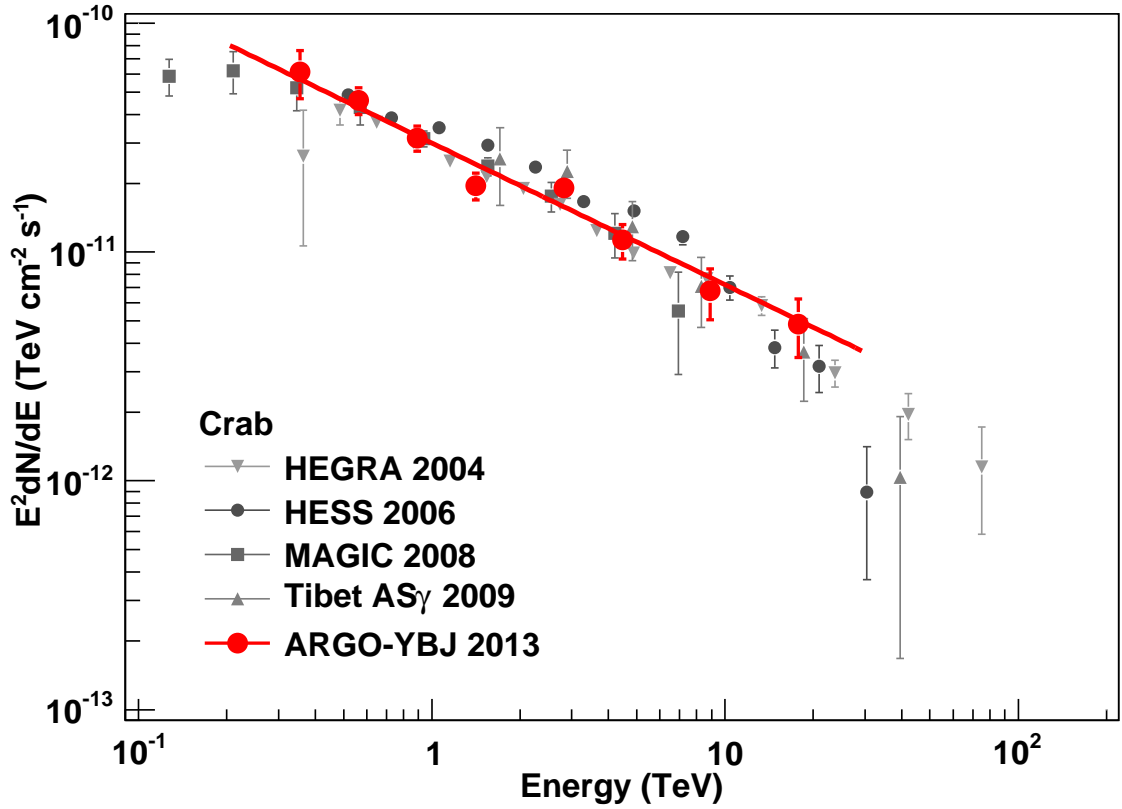}
\caption{Spectral energy distribution of the Crab Nebula   measured by ARGO-YBJ   and comparison with the measurements of   HEGRA, H.E.S.S., MAGIC,   and Tibet AS$\gamma$ \citep{aharon04,aharon06,albert08b,amenom09}. The solid line is the best fit to the ARGO-YBJ data using a power-law function. }
\label{fig5}
\end{figure}

\begin{figure}
\plotone{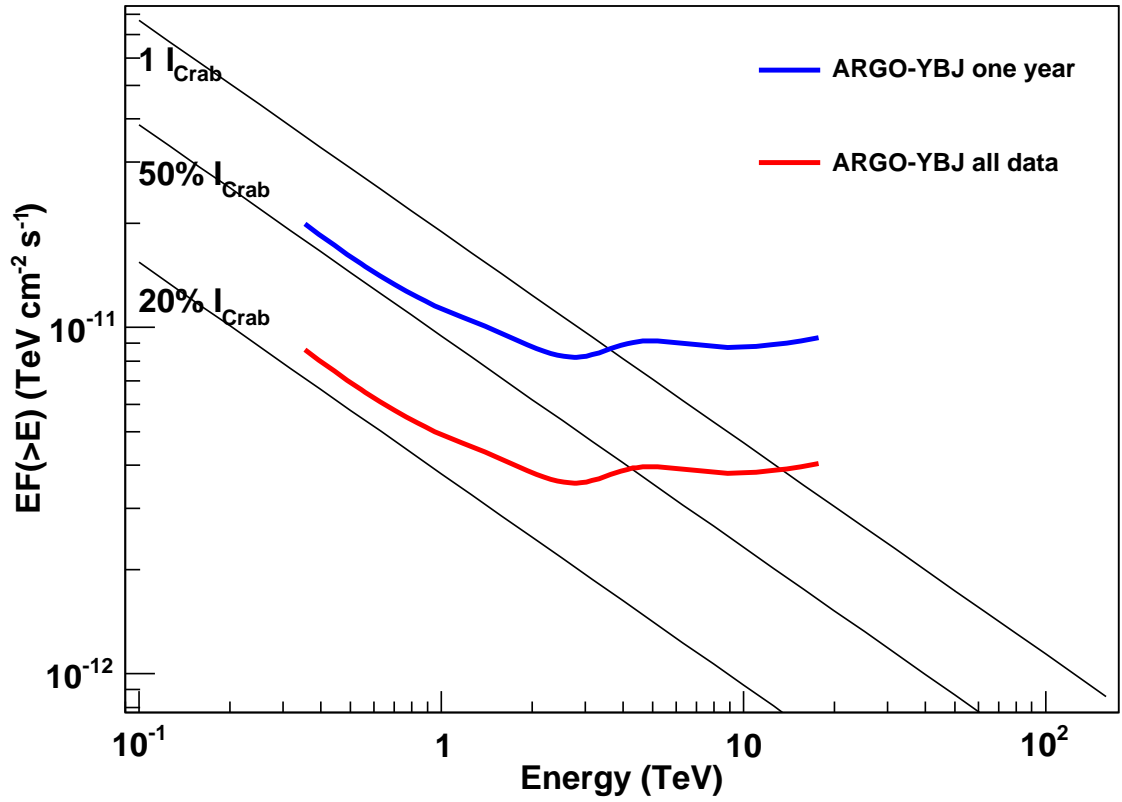}
\caption{Sensitivity curve of the ARGO-YBJ detector estimated using its observation results on the Crab Nebula. The integrated sensitivity curve is obtained using five years of ARGO-YBJ data. The one-year sensitivity curve is scaled from  this result.
The  duty cycle of the ARGO-YBJ detector has been taken into account.  }
\label{fig6}
\end{figure}

\begin{figure}
\plotone{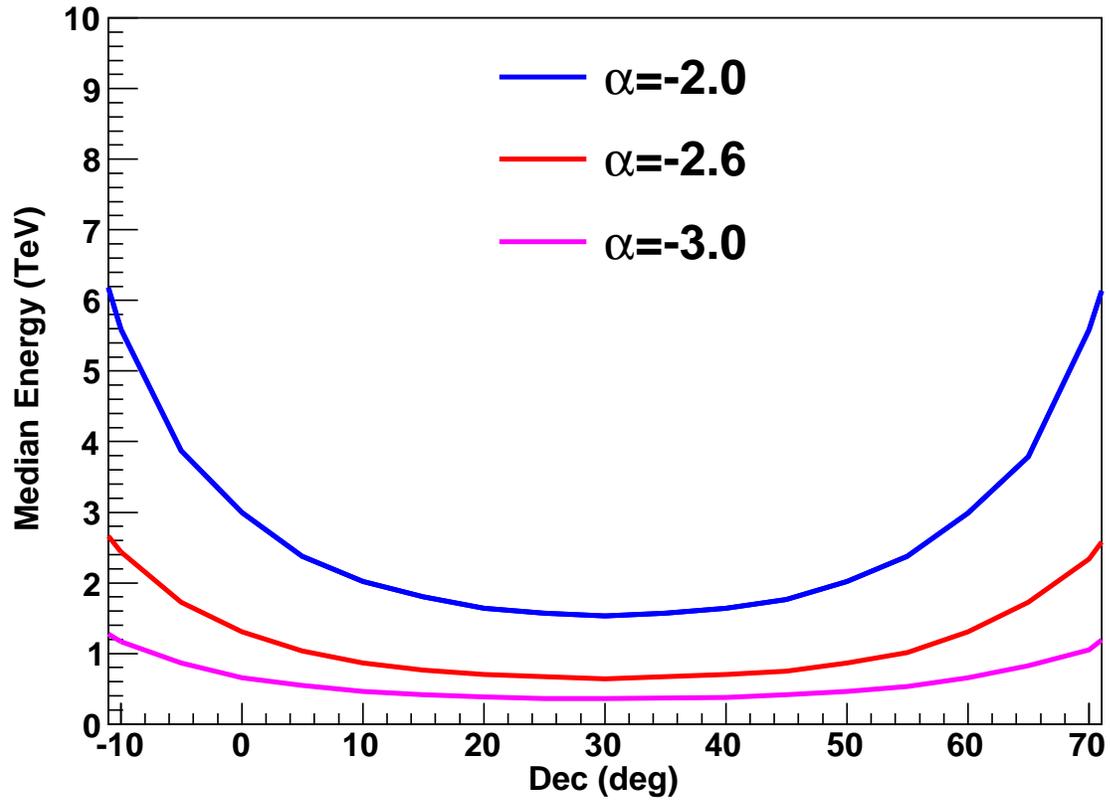}
\caption{Median energy of all the $\gamma$-ray events  that trigger ARGO-YBJ (N$_{pad}\geq20$) and satisfy the event selections  as a function of the source declination. Different lines correspond to different spectral indices, i.e., $-$2.0, $-$2.6, and $-$3.0.}
\label{fig7}
\end{figure}

\begin{figure}
\plotone{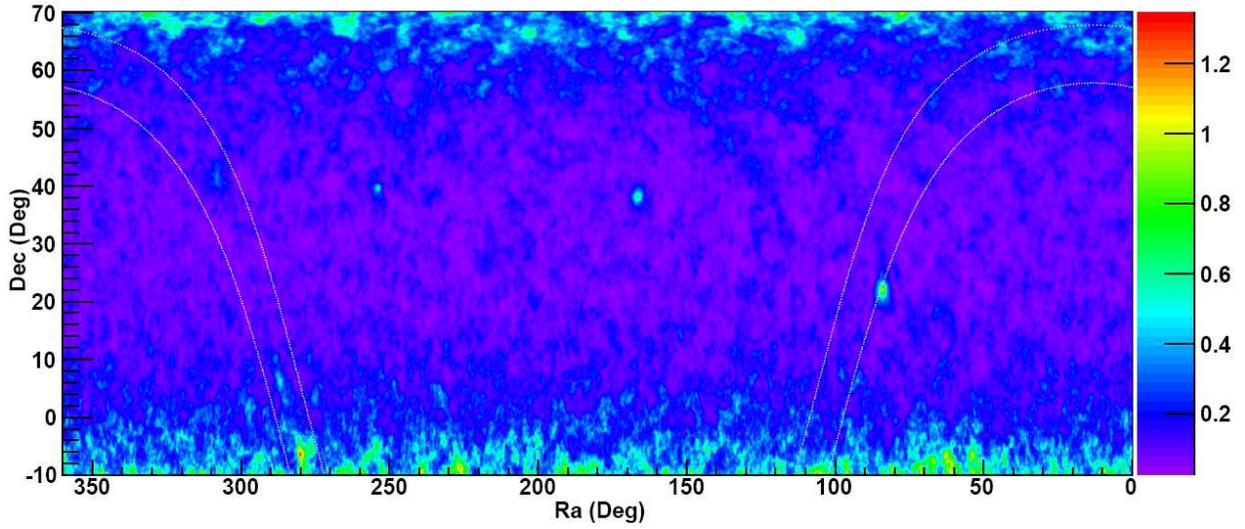}
\caption{Map of the 95\% C.L. flux upper limits at energies   above 500 GeV assuming an energy spectrum $E^{-2.6}$.
The color scale on the right is in Crab units, i.e., 5.69$\times 10^{-11}$ cm$^{-2}$ s$^{-1}$.
The two dotted lines indicate the Galactic latitudes $b=\pm$5$^{\circ}$.  }
\label{fig8}
\end{figure}

\begin{figure}
\plotone{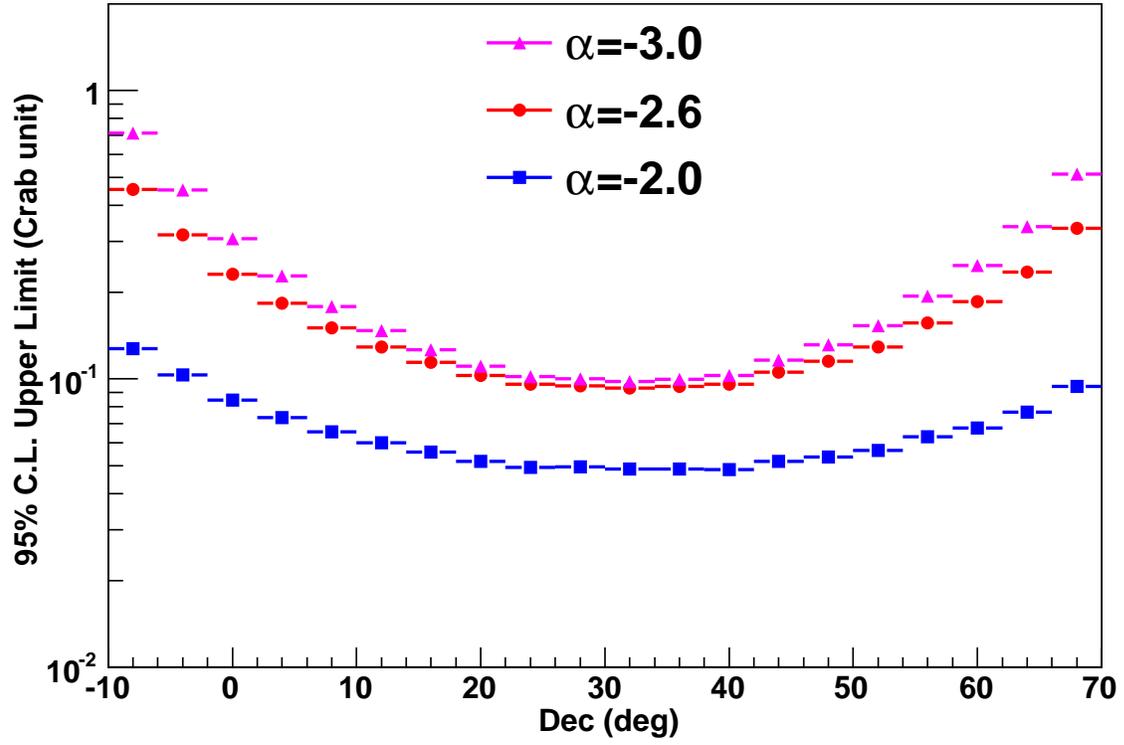}
\caption{   95\% C.L. flux upper limits at energies above 500 GeV  averaged on the right ascension, as a function of  the declination.   Different curves correspond to different power-law spectral indices  $-$2.0, $-$2.6 and $-$3.0. The Crab unit is 5.69$\times$10$^{-11}$ cm$^{-2}$ s$^{-1}$.}
\label{fig9}
\end{figure}

\begin{figure}
\plotone{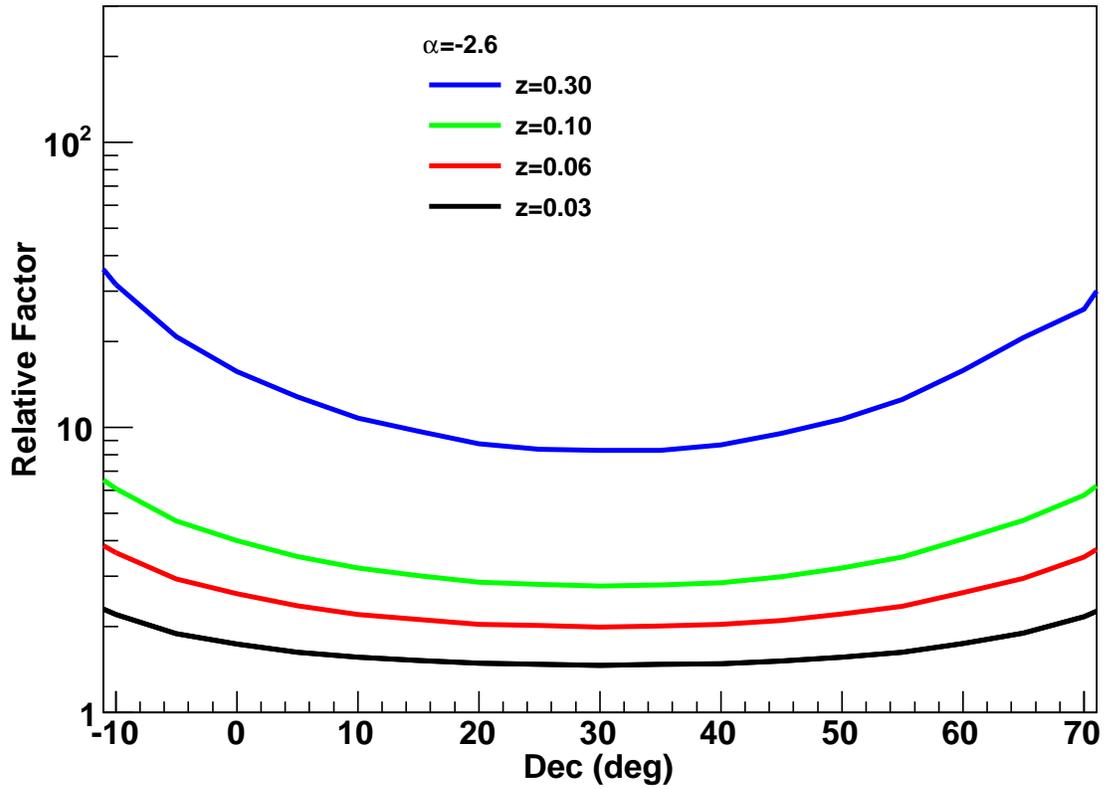}
\caption{ Effect of the EBL absorption on the upper limits shown in Figures 8 and 9. The source spectrum is assumed to be $E^{-2.6}$.
The $y$-axis gives the absorption factor for a source at the indicated redshift relative to a source at redshift zero  as a function of the declination.
}
\label{fig10}
\end{figure}

\begin{figure}
\epsscale{.90} \plotone{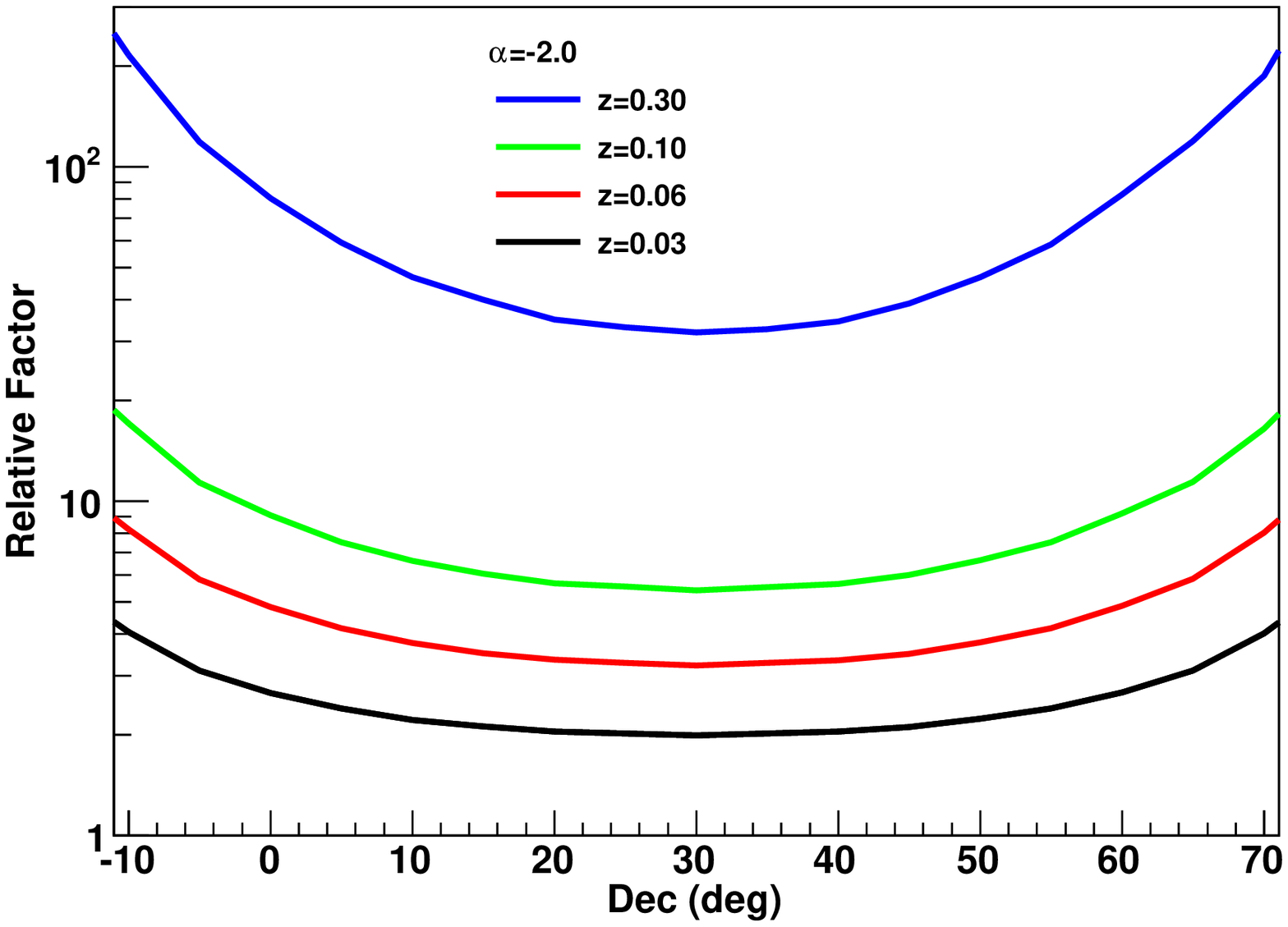}
\epsscale{.90} \plotone{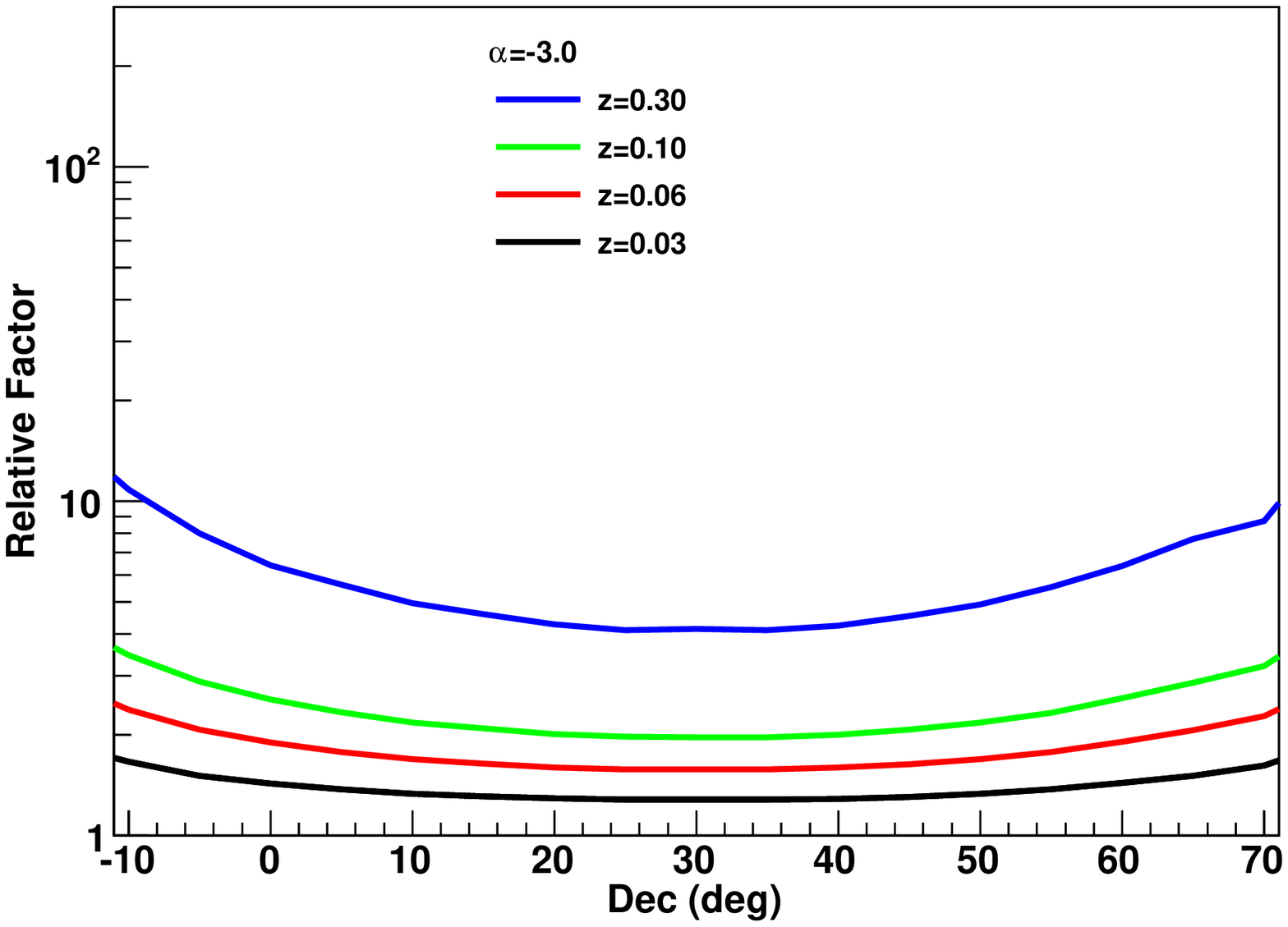}
\caption{ Effect of the EBL absorption on the upper limits shown in Figures 8 and 9. The source spectrum is assumed to be $E^{-2}$ (top panel) and $E^{-3}$ (bottom panel). The $y$-axis gives the absorption factor for a source at the indicated redshift relative to a source at redshift zero  as a function of the declination.
}
\label{fig11}
\end{figure}

\begin{figure}
\plotone{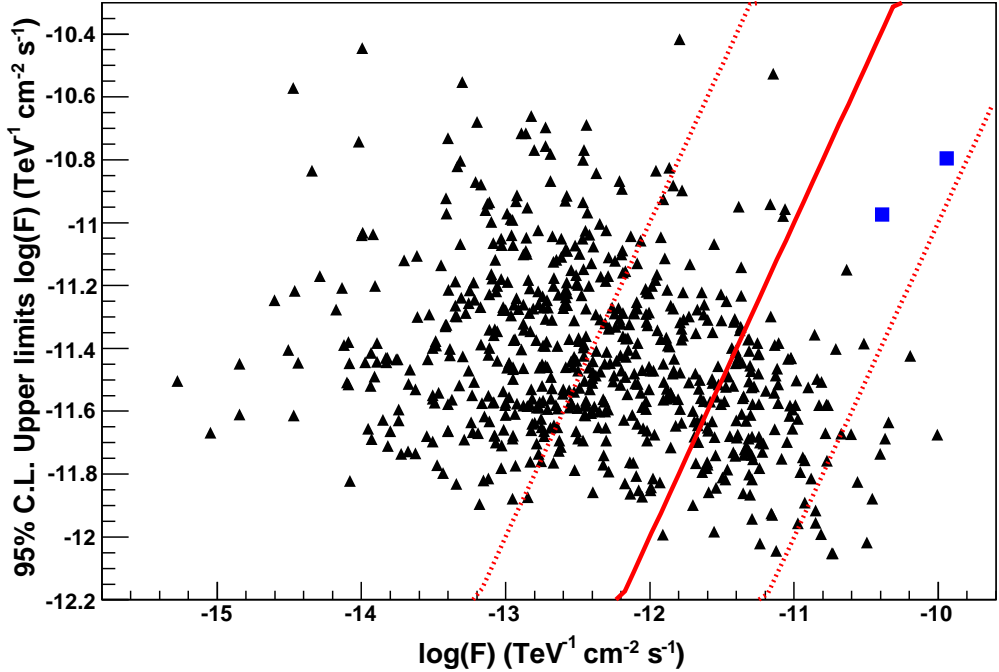}
\caption{
Comparison between ARGO-YBJ 95\% C.L. flux upper limits and the expected flux for the 663 $Fermi$-LAT AGNs   within the ARGO-YBJ FOV.
The expected fluxes are obtained by  extrapolating the SEDs  measured by $Fermi$-LAT   to TeV energies,   assuming  that the spectral index steepens by 0.5 at 100 GeV.
Both fluxes are   differential   at 1 TeV. The solid line indicates where the upper limit equals the expected flux. The dotted lines indicate the 0.1 and 10 times relations between these two fluxes. All the upper limits are estimated assuming the source at   redshift   zero. The two squares correspond to Mrk 421 and Mrk 501. }
\label{fig12}
\end{figure}

\begin{figure}
\plotone{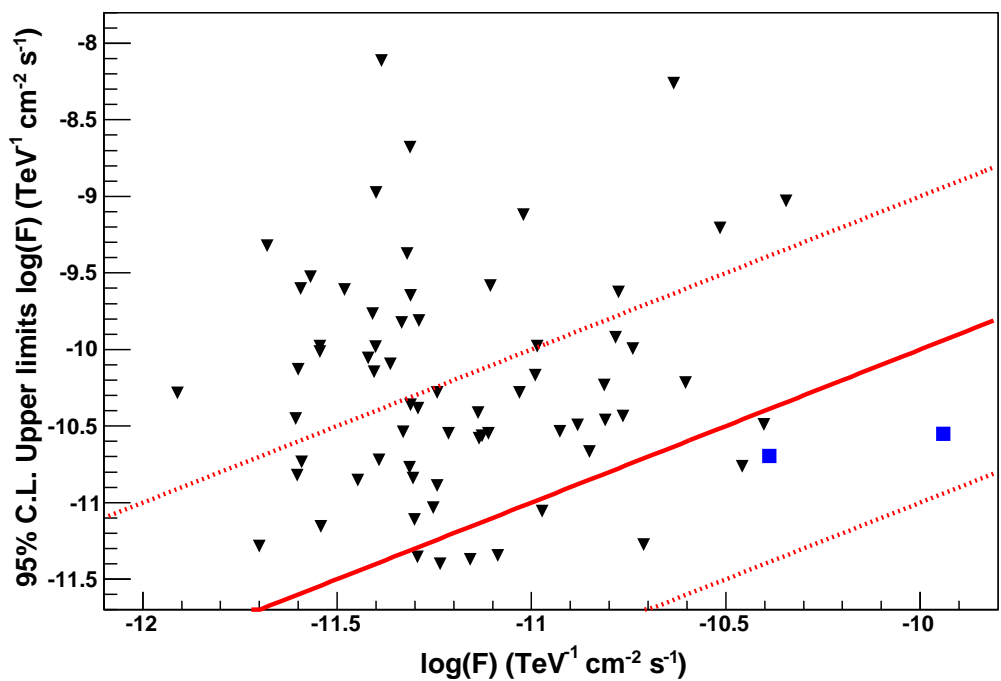}
\caption{ Comparison between ARGO-YBJ 95\% C.L. flux upper limits and the expected flux for   68 $Fermi$-LAT AGNs with measured redshift.  The expected fluxes are obtained by  extrapolating the SEDs  measured by $Fermi$-LAT   to TeV energies, assuming that the spectral index steepens by 0.5 at 100 GeV.  Both fluxes are   differential   at 1 TeV. The effect of the EBL absorption on the flux upper limits has been taken into account. The  lines represent the same flux relations as  in Figure 12. The two squares correspond to Mrk 421 and Mrk 501. }
\label{fig13}
\end{figure}
\clearpage


\begin{thebibliography}{}
\bibitem[Abdo et al.(2007)]{abdo07a}Abdo, A. A., et el. 2007, \apj , 664, L91
\bibitem[Abdo et al.(2009)]{abdo09}Abdo, A. A., et el. 2009, \apj, 700, L127
\bibitem[Ackermann et al.(2011a)]{acker11}Ackermann, M.,  Ajello, M.,  Allafort, A., et el. 2011, \apj, 743, 171
\bibitem[Ackermann et al.(2011b)]{acker11b}Ackermann, M.,  Ajello, M.,  Allafort, A., et el. 2011b, Science, 334, 1103
\bibitem[Aharonian et al.(2002)]{aharon02} Aharonian, F., et al. 2002, A\&A, 390, 39
\bibitem[Aharonian et al.(2004)]{aharon04}  Aharonian, F., et al. 2004, \apj, 614, 897
\bibitem[Aharonian et al.(2006)]{aharon06}  Aharonian, F., et al. 2006, A\&A, 457,   899
\bibitem[Aharonian et al.(2008a)]{aharon08}  Aharonian, F.,  et al. 2008a, A\&A, 477,   353
\bibitem[Aharonian et al.(2008b)]{aharon08c}  Aharonian, F.,  et al. 2008b, A\&A, 484,   435
\bibitem[Aharonian et al.(2008c)]{aharon08b} Aharonian, F.,  et al. 2008c, Rep. Prog. Phys. 71, 096901
\bibitem[Aharonian et al.(2009)]{aharon09}  Aharonian, F., Akhperjanian, A. G., Anton, G., et al. 2009, A\&A, 499, 723
\bibitem[Aielli et al.(2006)]{aielli06} Aielli, G.,  et al. 2006, Nucl. Instrum. Meth. A, 562, 92
\bibitem[Aielli et al.(2009a)]{aielli09} Aielli, G.,   et al. 2009a, Astropart. Phys., 30, 287
\bibitem[Aielli et al.(2009b)]{aielli09b} Aielli, G.,   et al. 2009b, Astropart.. Phys.,   32, 47
\bibitem[Aielli et al.(2009c)]{aielli09d} Aielli, G.,   et al. 2009c, \apj,   699, 1281
\bibitem[Aielli et al.(2009d)]{aielli09c} Aielli, G.,   et al. 2009d, Nucl. Instrum. Meth. A, 608, 246
\bibitem[Aielli et al.(2010)]{aielli10} Aielli, G.,   et al. 2010, \apj, 714, L208
\bibitem[Aielli et al.(2011)]{aielli11} Aielli, G.,   et al. 2011, \apj, 729, 113
\bibitem[Albert et al.(2008a)]{albert08} Albert, J., et al. 2008a, Science 320, 1752
\bibitem[Albert et al.(2008b)]{albert08b} Albert, J., et al. 2008b, \apj, 674, 1037
\bibitem[Aliu et al. (2013)]{aliu13} Aliu, E., et al. 2013, \apj, 770, 93
\bibitem[Atkins et al.(2004)]{atkins04} Atkins, R., et al. 2004, \apj, 608, 680
\bibitem[Amenomori et al.(2005)]{amenom05} Amenomori, M., et al. 2005, \apj, 633, 1005
\bibitem[Amenomori et al.(2009)]{amenom09} Amenomori, M., et al. 2009, \apj, 692, 61
\bibitem[Amenomori et al.(2010)]{amenom10} Amenomori, M., et al. 2010, \apj, 709, L6
\bibitem[Bartoli et al.(2011a)]{barto11} Bartoli, B.,   et al. 2011a, \apj, 734, 110
\bibitem[Bartoli et al.(2011b)]{barto11b} Bartoli, B.,   et al. 2011b, Phys. Rev. D, 84, 022003
\bibitem[Bartoli et al.(2012a)]{barto12a} Bartoli, B.,   et al. 2012a, \apj, 745, L22
\bibitem[Bartoli et al.(2012b)]{barto12b} Bartoli, B.,   et al. 2012b, \apj, 758, 2
\bibitem[Bartoli et al.(2012c)]{barto12c} Bartoli, B.,   et al. 2012c, \apj, 760, 110
\bibitem[Bartoli et al.(2013)]{barto13} Bartoli, B.,   et al. 2013, \apj, 767, 99
\bibitem[Capdevielle et al.(1992)]{capde92} Capdevielle, J.N., et al., 1992, KfK Report No.4998
\bibitem[Chen (2013)]{chen13} Chen, S.Z., 2013, Sci China-Phys Mech Astron, 56, 1454
\bibitem[Eckmann et al.(1991)]{eckma91} Eckmann, R., et al., 1991, in Proc. 22nd ICRC,  Dublin, Ireland, 4:464.
\bibitem[Fleysher et al.(2004)]{fleysher04} Fleysher, R., Fleysher, L., Nemethy, P.,  \&  Mincer, A.I., 2004, \apj, 603, 355
\bibitem[Franceschini et al.(2008)]{franc08} Franceschini, A., et al. 2008, A\&A, 487, 837
\bibitem[Gast et al.(2011)]{gast11}Gast, H.,  Brun, F.,  Carrigan, S., et al. 2011 ICRC, Beijing, 7, 158
\bibitem[Guo et al.(2010)]{guo}Guo, Y. Q., et al. 2010, CPC (HEP \& NP), 34(5):555
\bibitem[Hartman et al.(1999)]{hartman99} Hartman, R. C., et al. 1999, \apjs, 123, 79
\bibitem[He et al.(2007)]{he07} He, H. H., Bernardini, P., Calabrese Melcarne, A. K., \& Chen, S. Z. 2007, Astropart. Phys., 27, 528
\bibitem[Helene  (1983)]{helen83} Helene, O. 1983, Nucl. Instrum. Methods Phys. Res., 212, 319
\bibitem[Li \& Ma(1983)]{li83} Li, T.P.,\& Ma, Y.Q. 1983, \apj, 272, 317
\bibitem[Nolan et al.(2012)]{nolan12} Nolan, P. L., et al. 2012, \apjs, 199, 31
\bibitem[Pletsch et al.(2012)]{plets12} Pletsch, H.J., et al. 2012, \apj, 755, L22
\bibitem[Tueller et al.(2010)]{tuell10} Tueller, J.,   Baumgartner, W.H.,    Markwardt, C.B., et al. 2010, \apjs, 186, 378



\end{thebibliography}
\end{document}